\documentclass[sigconf]{acmart}
\AtBeginDocument{%
  }

\copyrightyear{2026}
\acmYear{2026}
\setcopyright{cc}
\setcctype{by}

\acmConference[DAC '26]{63rd ACM/IEEE Design Automation Conference}{July 26--29, 2026}{Long Beach, CA, USA}
\acmBooktitle{63rd ACM/IEEE Design Automation Conference (DAC '26), July 26--29, 2026, Long Beach, CA, USA}
\acmDOI{10.1145/3770743.3805813}
\acmISBN{979-8-4007-2254-7/2026/07}

\begin{document}

\title[]{DSPE: An Energy-Efficient Edge Processor for DeepSeek Inference with MerkleTree-based Incremental Pruning, Multi-Stage Boothing Lookup and Dynamic Adaptive Posit Processing}



\author{Yuhan Zhang}
\authornotemark[2]
\affiliation{%
  \department{School of Computer Science and Engineering}
  \institution{Northeastern University}
  \city{Shenyang}
  \country{China}
}

\author{Zhou Wang}
\authornote{Corresponding author.\\
$^\dagger$These authors contributed equally to this work.
}
\authornotemark[2]
\affiliation{%
  \institution{Imperial College London}
  \city{London}
  \country{United Kingdom}
}
\affiliation{%
  \institution{Imperial Global Singapore}
  \city{Singapore}
  \country{Singapore}
}

\author{Zhou Shu}
\author{Jiuren Zhou}
\affiliation{%
  \department{School of Microelectronics}
  \institution{Xidian University}
  \city{Xi'an}
  \country{China}
}
\affiliation{%
  \institution{Hangzhou Institute of Technology, Xidian University}
  \city{Hangzhou}
  \country{China}
}

\author{Yanqing Xu}
\affiliation{%
  \institution{The Chinese University of Hong Kong, Shenzhen}
  \city{Shenzhen}
  \country{China}
}

\author{Xiaonan Tang}
\affiliation{%
  \institution{Wisemaytech Co., Ltd.}
  \city{Beijing}
  \country{China}
}

\author{Shushan Qiao}
\author{Tianchun Ye}
\affiliation{%
  \institution{Institute of Microelectronics, Chinese Academy of Sciences}
  \city{Beijing}
  \country{China}
}
\affiliation{%
  \institution{University of Chinese Academy of Sciences}
  \city{Beijing}
  \country{China}
}

\author{Yang Liu}
\affiliation{%
  \institution{Nanyang Technological University}
  \city{Singapore}
  \country{Singapore}
}
\affiliation{%
  \institution{Imperial Global Singapore}
  \city{Singapore}
  \country{Singapore}
}

\author{Anil A. Bharath}
\author{Emm Mic Drakakis}
\affiliation{%
  \institution{Imperial College London}
  \city{London}
  \country{United Kingdom}
}
\affiliation{%
  \institution{Imperial Global Singapore}
  \city{Singapore}
  \country{Singapore}
}

\renewcommand{\shortauthors}{}

\begin{abstract}
In recent years, DeepSeek has achieved strong inference performance but remains hard to deploy on energy-constrained edge devices. This paper presents the DeepSeek Processing Element (DSPE), an edge-oriented architecture that alleviates the model’s heavy computational and energy demands. DSPE introduces three techniques: the MerkleTree-based Incremental Pruning Scheme (MIPS) for secure redundant-vector reduction, the Multi-Stage Boothing Lookup Method (MBLM) for bit-flip–aware approximate multiplication, and the Dynamic Adaptive Posit Processing Mechanism (DAPPM), which introduces a new DA-Posit format and its corresponding hardware multiplication architecture. Implemented in TSMC 28nm CMOS, DSPE achieves 109.4 TFLOPS/W energy efficiency compared with state-of-the-art designs and offers a scalable foundation for edge deployment.
\end{abstract}



\maketitle

\section{Introduction}
The rise of large language models (LLMs) such as BERT and GPT has driven significant advancements in natural language understanding, generation, and inference \cite{1,2,3,4}. The self-attention mechanism of Transformers makes long-sequence modeling more efficient, but as models scale up to hundreds of billions of parameters, the resulting substantial computational costs and massive storage requirements have become increasingly prominent. Against this backdrop, DeepSeek-V3 has attracted widespread attention for its outstanding inference efficiency and generalization capability. As shown in Figure\ref{f1}, the inference pipeline of DeepSeek-V3 consists of multiple stacked Transformer layers: each layer first maps input tokens into query (Q), key (K), and value (V) vectors \cite{5,6,7,8}, which are then processed collaboratively in the attention sublayer and the Multilayer Perceptron (MLP) sublayer. To further enhance representation capability, the attention component uses a multi-head structure to capture features from different semantic subspaces in parallel; in the MLP sublayer, gated activation and expert selection mechanisms allow different tokens to be routed to the most appropriate expert modules for nonlinear transformations as needed \cite{9,10,11}. Meanwhile, DeepSeek-V3 maintains a KV Cache during inference, enabling the reuse of historical information in long-sequence scenarios and effectively reducing redundant computation \cite{12,13,14}.

\begin{center}
    \includegraphics[width=0.8\linewidth]{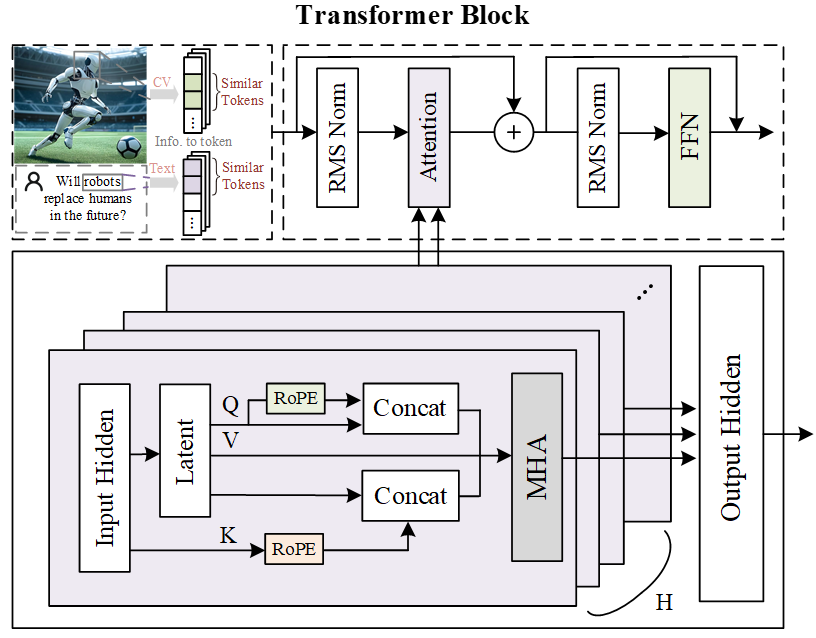}  
    \captionof{figure}{Diagram of Deepseek-V3.}
    \label{f1} 
\end{center}

In the development of large-scale language models, as parameter redundancy becomes increasingly significant, both academia and industry have proposed various optimization directions to improve inference efficiency \cite{15}. On one hand, algorithm-level approaches include model distillation, low-rank matrix factorization, sparsification, and MoE, which enhance inference performance by reducing the number of parameters or lowering computational complexity \cite{16}. On the other hand, at the hardware level, accelerators such as GPUs and TPUs use systolic arrays and mixed-precision techniques to improve throughput while maintaining model size. These methods have achieved certain success in the deployment of general LLMs, but they still struggle to effectively address the computational redundancy between parameters \cite{17,18,19,20}. Meanwhile, for models like DeepSeek-V3, which feature ultra-large-scale parameters and distributed MoE characteristics, dedicated architectural support is still lacking \cite{21}.

To overcome the above limitations, this paper proposes a high-efficiency processor architecture specifically designed for distributed inference of DeepSeek, called the DeepSeek Processing Engine (DSPE). Unlike conventional solutions that rely on static sparsity or uniform operand assumptions, DSPE can adaptively adjust the computation process in real time based on input data characteristics, layer types, and distributed workloads. The main contributions of this paper are summarized as follows:
\begin{enumerate}
    \item \textbf{MerkleTree-based Incremental Pruning Scheme} \\
    We propose the MerkleTree-based Incremental Pruning Scheme (MIPS), which alleviates computational redundancy in edge inference while enhancing security through a three-stage pipeline: "similarity reordering—Merkle Tree encoding—dynamic reuse." The module maintains semantic continuity using a sequence incremental sorter and achieves secure and efficient skipping of computations and data reuse through Merkle tree encoding. Combined with the result reuse mechanism for corresponding similar features, MIRA improves inference energy efficiency while reducing bandwidth overhead.
    
    \item \textbf{Multi-Stage Boothing Lookup Method} \\
    We design the Multi-stage Booth Lookup Method (MBLM) to optimize multiplication operations in scenarios involving shared weights. The module introduces a Bayesian network to adaptively choose between radix-4 and radix-8 Booth encoding paths, dynamically balancing computational efficiency and energy consumption under different input distributions. Through partial product reordering and early termination mechanisms, it effectively reduces encoding redundancy and memory access operations.

    \item \textbf{Dynamic Adaptive Posit Processing Mechanism} \\
    We design the Dynamic Adaptive Posit Processing Mechanism (DAPPM), which includes the DA-Posit numerical system and its compatible hardware architecture. DA-Posit achieves three types of structured compression by merging and restructuring the identical parts of the lower-order fraction and exponent, and performs pattern decision and computation through branch paths. It reduces power consumption while maintaining precise numerical values, providing an efficient arithmetic foundation for edge DeepSeek inference.
    
\end{enumerate}

\section{PRELIMINARIES AND MOTIVATION}
\subsection{DeepSeek-V3}
DeepSeek-V3 is a large-scale Transformer optimized for long sequences and autoregressive inference, built based on Multi-Head Latent Attention (MLA) and sparse Mixture of Experts (MoE). The input tokens are embedded, combined with positional encodings, and then fed into stacked Transformer blocks. For single-head attention, its computation can be shown as:
\begin{equation}
\text{Attention}(Q,K,V) = \text{Softmax}\left(\frac{QK^T}{\sqrt{d_k}}\right)V
\label{eq:single_head_attention}
\end{equation}
And MLA can be shown as:
\begin{equation}
\text{MLA}(X) = \text{Concat}(\text{head}_1, \dots, \text{head}_h) W^O
\label{eq:mla}
\end{equation}
In MLA, the Query focuses on cached Key/Value pairs, allowing efficient reuse of historical context in long sequences and autoregressive settings. The feedforward stage can be implemented as a dense MLP or a sparse MoE layer, with the latter's computation taking the form of:
\begin{equation}
\text{MoE}(x) = \sum_{i=1}^E G_i(x) \text{FFN}_i(x)
\label{eq:moe}
\end{equation}
Here, $G_i(x)$ is a sparse gating function that assigns non-zero weights to only a few experts.

\subsection{Posit}
The Posit format is a compact, dynamically encoded floating-point representation, defined by the bit width ($n$) and exponent size ($es$), offering better precision range than floating-point numbers. The specific format consists of the following four parts:

1) Sign: 0 represents a positive number, and 1 represents a negative number.
    2) Regime: The scale factor is represented using a run-length encoding method. It consists of several consecutive identical bits (0 or 1), and ends when an opposite bit appears or the end of the string is reached.
    3) Exponent: Posit can directly reflect the magnitude of numerical changes.
    4) Fraction: If there are bits remaining after the exponent bits, they represent the fraction.

\subsection{Merkle Tree}
\begin{center}
    \centering
    \includegraphics[width=0.8\linewidth]{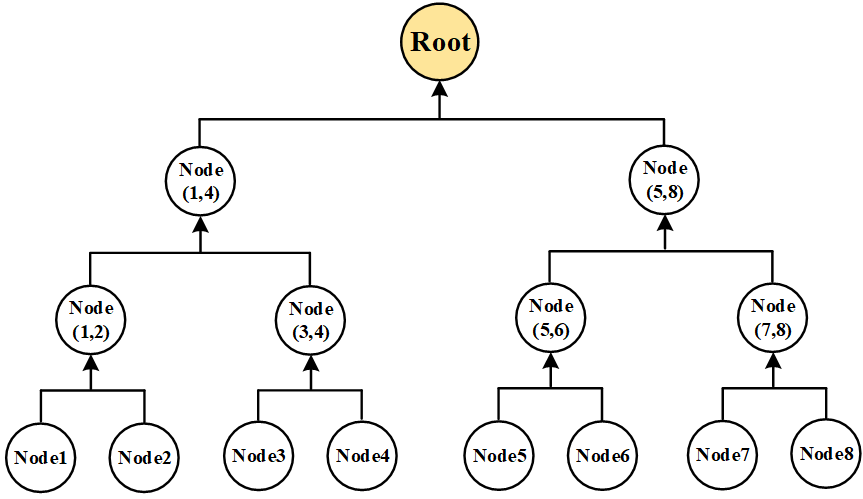} 
    \captionof{figure}{Merkle Tree Structure.}
    \label{f2}  
\end{center}

The Merkle tree is a hash-based hierarchical structure that efficiently checks integrity by recursively combining leaf hashes into a unique root, allowing the integrity and security of data to be verified through the consistency of the root. Another important value of this is that when the calculation result is uniquely determined, the calculated values of its intermediate nodes are also uniquely determined, which can be used for advance prediction of data.

\subsection{Challenges of DeepSeek-Base Processor}
As shown in Figure\ref{f3} , the challenges of deploying DeepSeek on edge devices are related to the characteristics of its hierarchical inference process and can be mainly categorized into three types:
\begin{figure}[ht]
    \centering
    \includegraphics[width=0.8\linewidth]{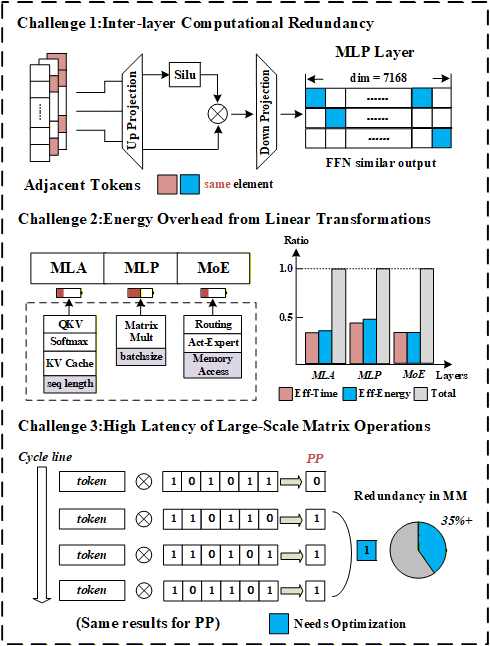} 
    \caption{Challenges of DeepSeek Inference on Edge Devices.}
    \label{f3}  
\end{figure}
\begin{enumerate}
    \item \textbf{Inter-layer Computational Redundancy}: \\
    DeepSeek generates a large number of spatiotemporally redundant intermediate representations during its phased processing. These intermediate representations have little impact on the final result (e.g., similar transformations of adjacent terms, overlapping key-value cache contexts, and repeated computations by the same expert). On edge devices, this manifests as resource waste, but it also provides an opportunity to achieve high-efficiency optimization.
    
    \item \textbf{Energy Overhead from Linear Transformations}: \\
    In DeepSeek, the extensive repeated large-scale linear transformations, attention computations, and routing decisions in the hierarchical process (especially dense MLPs, sparse MoEs, and the maintenance and access of KV caches in MLAs) can cause significant signal flipping and memory power consumption, making it easy to reach or exceed the power limits on energy-constrained edge platforms.
    
    \item \textbf{High Latency of Large-Scale Matrix Operations}: \\
    In inference scenarios, matrix multiplication is gradually adopting the Booth multiplication scheme to reduce partial products and accelerate computation. However, when the same weight is multiplied by multiple sets of activations and there are significant differences in their Booth encoding, it can cause a large number of bit flips and significantly increase energy consumption.
\end{enumerate}

\section{PROPOSED DSPE PROCESSOR}
\begin{figure}[ht]
    \centering
    \includegraphics[width=\linewidth]{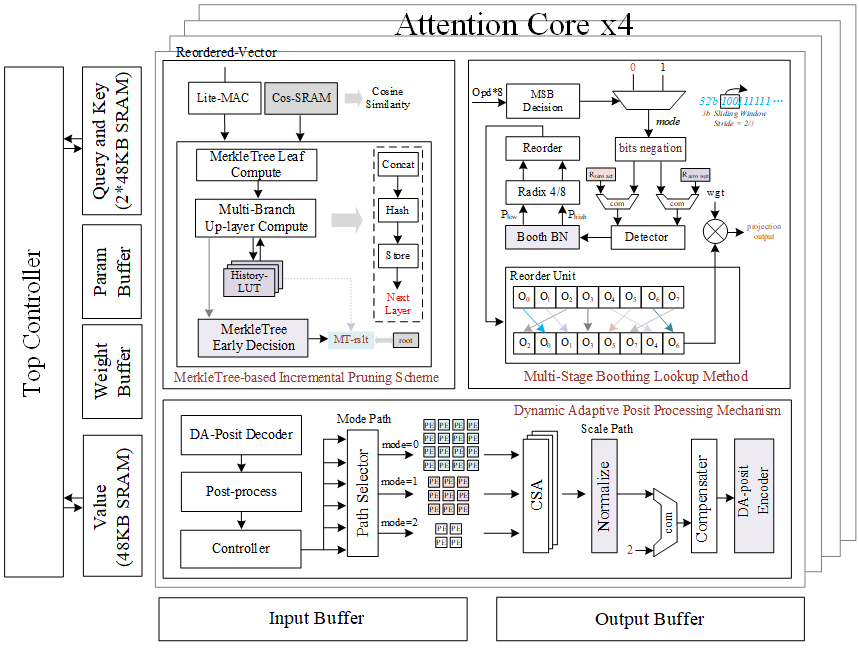} 
    \caption{The Architecture of the DeepSeek Processing Element.}
    \label{f4}  
\end{figure}

The overall architecture of DeepSeek Processing Element (DSPE) processor includes 4 Attention Cores (ACs), as shown in Figure\ref{f4}, which uses three core technologies: the Merkle-Based Incremental Pruning Scheme, the Multi-stage Boothing Lookup Method, and the Dynamic Adaptive Posit Processing Mechanism. Each AC's data path contains an iRouter and an oRouter. The iRouter dynamically distributes activations and weights to the target processing elements (PEs). Additionally, the architecture includes an Input Buffer and an Output Buffer for caching input data and computation results, respectively. A 24KB Parameter Buffer and a 48KB Weight Buffer are used to pre-store intermediate parameters and reusable weight data. The Top Controller centrally schedules data flow and memory access: on one hand, it controls two 48KB Query/Key dedicated SRAMs and manages one 48KB Value dedicated SRAM. On the other hand, the Parameter Buffer can load parameter blocks of different layers according to the scheduling strategy of the Top Controller, enabling pipelined weight reuse; while the Weight Buffer works in coordination with the iRouter to ensure efficient broadcasting and sharing of weight tensors between different Attention Cores.

\subsection{MerkleTree-based Incremental Pruning Scheme (MIPS)}
\begin{figure}[t]
    \centering
    \includegraphics[width=\linewidth]{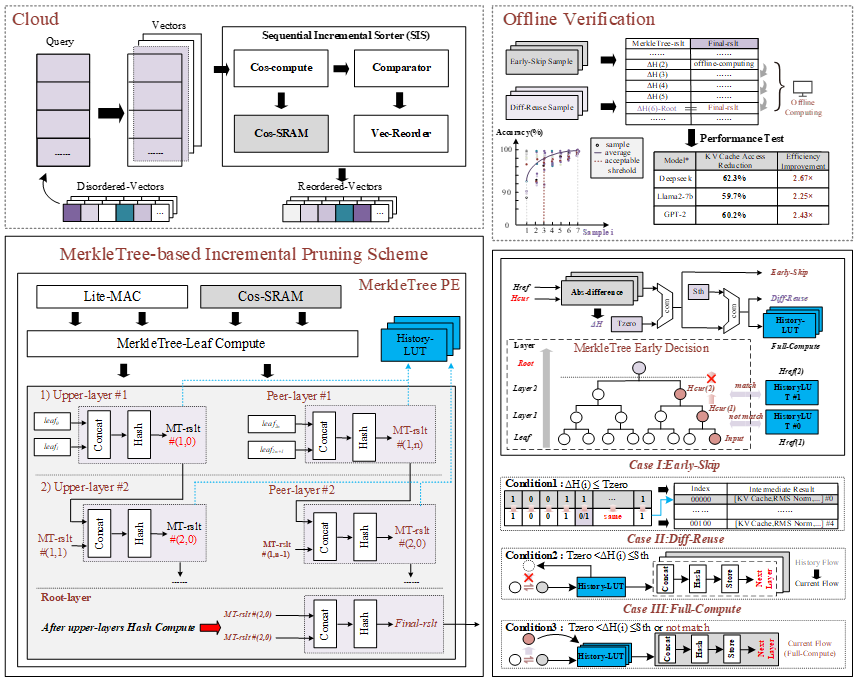} 
    \caption{MerkleTree-based Incremental Pruning Scheme.}
    \label{f5}  
\end{figure}

In the token-level decoding phase of DeepSeek, each new query attends to an ever-growing KV cache, while adjacent tokens produce highly similar Q/K vectors, causing redundant similarity evaluations and KV accesses. To exploit this redundancy, we design a MerkleTree-based Incremental Pruning Scheme (MIPS), as shown in Figure\ref{f5}, which uses a three-stage pipeline of similarity reordering, MerkleTree early decision, and dynamic reuse.

Firstly, the MIPS module introduces a Sequential Incremental Sorter at the software level: newly generated vectors are inserted into the appropriate positions in incremental order based on their cosine similarity with vectors already present in the current sequence, while all cosine similarity scores are cached in Cos-SRAM. Secondly, the reordered vectors are fed into the MerkleTree PE. The high-dimensional vectors are projected onto a compact semantic space using a lightweight MAC operation, resulting in $V_{\text{low}} = \text{MAC}(V_{\text{reordered}})$. Then, for each new input vector, the MerkleTree PE computes leaf nodes using locality-sensitive hashing by reusing cosine similarity from Cos-SRAM, progressively building the Merkle tree from bottom to top. At each layer, such as the $i$-th layer, the hash value of the current intermediate node is recorded as $H_{\text{cur}}(i)$, and compared with the hash value $H_{\text{ref},j}(i)$ of the intermediate node at the same layer in the historical input processed by the same expert. The difference is calculated as: 
$
\Delta H_j(i) = |H_{\text{cur}}(i) - H_{\text{ref},j}(i)|
$
Two thresholds, $T_{\text{zero}}$ and $S_{\text{th}}$, are respectively set to determine whether redundant or approximate computations exist.

MIPS can make three types of decisions at the $i$-th layer:
1) \textbf{Early-Skip}: If $\Delta H_j(i) \leq T_{\text{zero}}$, it is considered that the current input is completely identical to the historical samples at this layer in the hash space or differs only at the bit-level 0/1. MIPS directly reuses the corresponding KV Cache index and intermediate results, and treats the current round as repeated computation and skips it.
   2) \textbf{Diff-Reuse}: If $T_{\text{zero}} < \Delta H_j(i) \leq S_{\text{th}}$, and $\Delta H_j(i)$ hits in $\text{History-LUT}_i$, it indicates that the differential mode for this layer has previously completed a Full-Compute and registered a reusable result. At this point, MIPS directly looks up the result index, terminates higher-level Merkle construction and subsequent computations, and switches the current pipeline to the reuse path.
  3) \textbf{Full-Compute}: If $\Delta H_j(l) > S_{\text{th}}$, or even if it falls within the approximate range but no matching entries are found in the entire History-LUT, the current input is considered a “completely new pattern”. MIPS continues to build higher-level nodes up to the root hash. After completion, $\Delta H$ and its final result are written back to History-LUT for future reuse and statistical purposes.
To evaluate the robustness of early judgments, MIPS reserves necessary statistical interfaces on the hardware side for the upper-level system. System software can, in an offline background mode, recompute the complete root hash for samples to statistically evaluate the consistency between various early decisions and the root node judgment. 

The MMLU dataset is often used to comprehensively evaluate the cross-domain knowledge understanding and reasoning capabilities of large language models. It is evaluated for DeepSeek-V3 in MMLU dataset. Through MMLU evaluation, we found that MIPS saves 33.5\% of DRAM memory access and 36.2\% of SRAM memory access by performing lookup reuse and Early Decision at intermediate nodes of the MerkleTree.

\subsection{Multi-Stage Boothing Lookup Method (MBLM)}
\begin{figure}[!t]
    \centering
    \includegraphics[width=\linewidth]{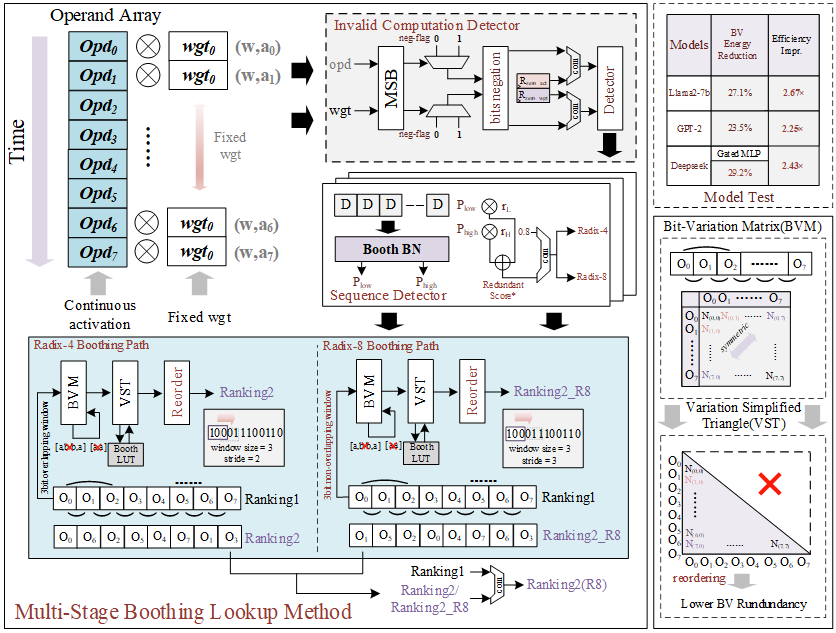}  
    \captionof{figure}{Multi-Stage Boothing Lookup Method.}
    \label{f6}
    \vspace*{-0.3cm}
\end{figure}

During the inference phase of DeepSeek, the matrix multiplication operations in the MLP layers exhibit obvious temporal locality: activations coming from adjacent time frames and sequentially entering the computing units are multiplied with a fixed set of weights, forming a computational pattern of “multiple multipliers × the same multiplicand.” However, Traditional matrix multiplication arrays lack awareness of reuse opportunities brought by temporal proximity. To address this, this paper proposes the Multi-Stage Boothing Lookup Method (MBLM), as shown in Figure\ref{f6}, introducing an invalid computation detector, a Bayesian Network (BN) model for redundant classification, and a branched boothing lookup pipeline, and is used to differentiate vector redundancy caused by spatiotemporal locality.

First, MBLM takes 8 multiplication operands into the invalid computation detector at a time. If the weight or activation values are very small, their contribution can be ignored. Based on this observation, the module sets near-zero weight and activation threshold $R_{\text{zero\_wgt}}$ and $R_{\text{zero\_act}}$. Multiplication pairs that meet $|w| < R_{\text{zero\_wgt}}$ or $|a| < R_{\text{zero\_act}}$ are directly marked as invalid computations and skipped. On this basis, MBLM introduces a Booth BN model inside the sequence detector unit to model and classify the similarity between operands: by extracting the bit similarity (BS) of adjacent token multiplication requests and the length of repeated fragments (Repeat Length, Re-length), a redundant feature representation of the input sequence is obtained. BS is shown in Equation 4,
$\text{BV}$ refers to the number of bit-flips between adjacent multiplication requests.
\begin{equation}
\text{BS} = 1 - \frac{\text{BV}}{8}
\label{eq:bit_similarity}
\end{equation}

BN outputs two probabilities, $P(R=\text{Low})$ and $P(R=\text{High})$, and calculates the redundancy score based on them, which is shown in Equation 5:
\begin{equation}
\text{redundant score} = r_L \cdot P(R=\text{Low}) + r_H \cdot P(R=\text{High})
\label{eq:redundancy_score}
\end{equation}
When the redundancy score is below 0.8, MBLM chooses the radix-4 Booth as the regular path; when the score is above 0.8, it switches to the radix-8 Booth extended path, further reducing the number of partial products under higher radix encoding. The module accumulates statistics on BV between any two multiplicands to construct an 8×8 Bit-Variation Matrix (BVM), and removes two types of redundancy: Case I is duplicate counting of exchange pairs like “A and B” and “B and A”; Case II is invalid counting such as “A and A”. This allows the effective statistical area to be simplified into a Variation Simplified Triangle (VST). The sorting result output by the regular path is denoted as ranking2, and the result output by the extended path is denoted as ranking2\_R8. 

MBLM further incorporates a lightweight repetition detection and lookup reuse mechanism. Each VST is equipped with a small Booth-LUT that records the BV flip pattern and corresponding sequential index from the last execution on that entry. When a new multiplication pair maps to a particular VST entry, the repetition detection unit first compares its BV flip distribution with the table entry; if the BV count is zero or below the preset "complete match" threshold, it is considered a fully redundant computation at the current precision, and Booth encoding and partial product generation are skipped. The results are finally compared by a comparator, which selects the execution order.

MBLM-based analysis reveals that the amount of calculation can be reduced by 39.1\% on DeepSeek-V3 in MMLU dataset, which means the speed of calculation can be further improved. The significantly reduced bit flipping in the rearranged sequence after Booth encoding in MBLM achieves redundancy reduction. 
\begin{table*}[h]
    \centering
    \caption{A COMPARISON OF PROCESSOR DESIGNS}
    \label{t4}  
    \resizebox{\linewidth}{!}{
    \begin{tabular}{lcccccc}
        \toprule
        & GPU H100{[}1{]} & ISSCC'23 {[}6{]} & ISSCC'23 {[}7{]} & VLSI'24{[}8{]} & This Work \\
        \midrule
        Training Support & YES & NO & NO & YES & YES \\
        Technique (nm) & 4 & 12 & 28 & 22 & 28 \\
        Die Area (mm$^2$) & 814 & 4.6 & 14.36 & 6.4 & 8.23 \\
        Supply Voltage (V) & NA & 0.62 - 1.0 & 0.6 - 1.0 & 0.6 - 1.0 & 0.6 - 1.1 \\
        Frequency (MHz) & 1620 & 77 - 717 & 85 - 275 & 115 - 495 & 200 - 710 \\
        Precision & FP64/32/8 \newline INT32/8/4 & FP8/4 & INT16/8 & BF16/FP8 & POSIT8, INT8/4 \\
        Power (mW) & 700000 & 10 -122@FP8 \newline 9 - 111 @FP4 & 29.83 - 152.75 & 49.2 - 451.2 & 122 - 345 \\
        Performance (TOPS or TFLOPS) & 1978.9 @FP16 \newline 3957.8 @FP8 & 0.367 @FP8 \newline 0.734 @FP4 & 0.89 @INT16 \newline 3.55 @INT8 & 2.38 @BF16 \newline 5.69 @FP8 & 22.8 TFLOPS @POSIT8 \\
        Energy Efficiency (TOPS/W or TFLOPS/W) & 2.827 @FP16 \newline 5.654 @FP8 & 8.24 @FP8 \newline 18.1 @FP4 & 60.3 @INT16 \newline 101.1 @INT8 & 20.58@BF16 \newline 54.94 @FP8 & 109.4 TFLOPS/W@POSIT8 \\
        \bottomrule
    \end{tabular}
    }
  
\end{table*}
\subsection{Dynamic Adaptive Posit Processing Mechanism (DAPPM)}
In Deepseek edge inference scenarios, significant hardware overhead is often caused by the coexistence of multiple precisions and precision conversion. The Posit numerical system is used to replace floating-point numbers due to its flexibility and wide range. However, typical inference-friendly Posits often adopt a format where the exponent width is the same as or close to the fraction width. In practice, the exponent and fraction boundaries are not always fully utilized, causing some bits to remain in a redundant state that carries little information for long periods.

To address this, this paper proposes a unified representation format for DSPE called Dynamic Adaptive Posit (DA-Posit). DA-Posit leverages the regime, which encodes an effective exponent range, and treats the exponent and fraction as a reconfigurable "dynamic precision field" (Dyn-field). When low-order bits are highly correlated or redundant, structured merging compression is applied: for example, with an 8-bit Posit, when the last 2 or 1 bits of the exponent and fraction are the same, they are folded into shared bits. Even in ultra-low precision configurations, an additional bit can be saved via end-bit folding. These operations only affect the low-order bits, which have minimal impact on numerical values, and the original Posit semantics can be losslessly restored using the scale information provided by the regime. At the same time, DA-Posit reuses a very small number of boundary regime codes, extending them into a "scale + mode" joint mapping to distinguish three modes— “no / 1-bit / 2-bit compression”—without increasing any encoding bit width, achieving an adaptive Posit representation with zero additional bit overhead.

\begin{figure}[h]
    \centering
    \includegraphics[width=\linewidth]{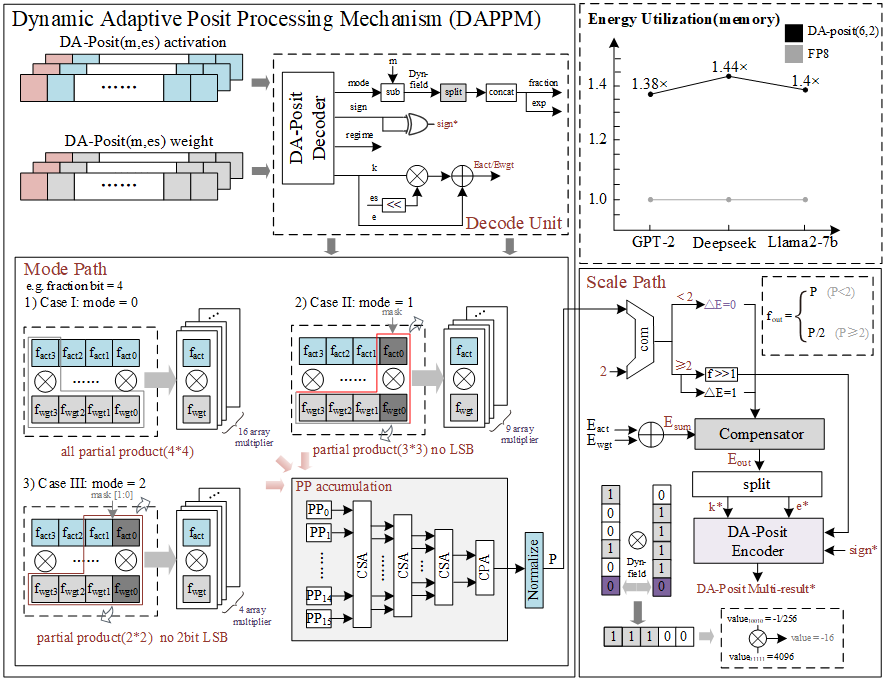} 
    \caption{Dynamic Adaptive Posit Processing Mechanism.}
    \label{f7} 
\end{figure}

For DA-Posit data, this paper proposes a dynamic adaptive Posit processing mechanism, as shown in Figure\ref{f7}. First, the input is decoded into sign, regime, $(k, e)$, and Dyn-field using a DA-Posit decoder, and $(k, e)$ is synthesized into a composite exponent $E$ using equation (6): 
\begin{equation}
E = k \cdot 2^{\text{es}} + e
\label{eq:composite_exponent}
\end{equation}
Then, the mode path is entered: depending on the mode value (0/1/2), a calculation configuration of 16, 9, or 4 Array Multiplier PEs is selected. The resulting partial product undergoes multi-level CSA Tree and CPA operations in the PP Accumulation Unit to obtain the Posit multiplication result and complete normalization. Next, the data enters the scale path. First, it checks whether the normalization result falls within the preset range $(0, 2)$. If it does not, compensation and correction are performed. If it does, the multiplication result is encoded using DA-Posit encoding, and then restored to DA-Posit format.

For DeepSeek-V3 on the MMLU dataset, the DAPPM method improves computation speed by 1.47 times while maintaining the same computational accuracy through three bit compression mechanisms of DA-Posit and flexible Array Multiplier PE mode selection.

\section{EVALUATION RESULTS}
\begin{figure}[t]
    \centering
    \vspace*{-0.1cm}
    \includegraphics[width=0.55\linewidth]{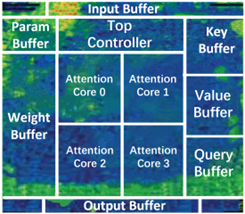}  
    \caption{The layout of the proposed processor.}
    \label{f8}  
    \vspace*{-0.1cm}
\end{figure}

Based on the proposed DSPE approach, we used Verilog to implement a DeepSeek inference processor. In order to achieve accurate evaluation, we synthesized the processor by means of the Synopsys Design Compiler tool, and employed a 28nm CMOS technology to perform placement and routing via the IC Compiler. Figure\ref{f8} depicts the layout of the implemented processor. It contains 4 PE arrays or Attention Cores, with each PE array containing 64 PE cores. The total area consumed is 8.23 mm$^2$. The proposed processor's supply levels range from 0.6V to 1.10V, while its power consumption ranges from 122mW to 345mW. The processor attains the maximum frequency of 710MHz when operating at 1.10V, and achieves 22.8TFLOPS@POSIT8 with the highest performance. When working at 0.6V and 200M frequency, the peak energy efficiency of the processor reaches 109.4TFLOPS/W@POSIT8. Comparative results with other processors are tabulated in Table\ref{t4}.

\section{Conclusion}
DeepSeek has demonstrated outstanding performance in large-scale language model inference, but its high computational demands and energy consumption have limited edge-side deployment. This paper proposes the DeepSeek Processing Element (DSPE)—a high-efficiency edge inference processor architecture designed for the DeepSeek model. This paper introduces three architectural innovations: First, the MerkleTree-based Incremental Pruning Scheme (MIPS), which reduces redundant vector computation and bandwidth usage through a three-stage pipeline of similarity reordering, MerkleTree-based early decision, and dynamic reuse, enhancing security while minimizing memory consumption; second, the Multi-Stage Boothing Lookup Method (MBLM) enables joint weight–activation approximate computing through a redundancy-score-based classifier that guides operand reordering, adaptive Booth encoding, and table replay, outperforming static weight-only pruning in reducing bit-flip energy and redundant computations; third, the Dynamic Adaptive Posit Processing Mechanism (DAPPM), featuring a novel dynamic Posit numerical system—DA-Posit and its hardware architecture, reducing data overhead and improving computational speed. This design is implemented using TSMC 28nm CMOS technology, achieving 109.4 TFLOPS/W peak energy efficiency, which is 19.35$\times$ higher than the GPU-H100@FP8.
\section*{Acknowledgments}
This research is supported by the National Research Foundation, Prime Minister’s Office, Singapore under its IN-CYPHER Campus for Research Excellence and Technological Enterprise (CREATE) Programme.
\newpage
\clearpage
\bibliographystyle{ACM-Reference-Format}
\bibliography{sample-base}
\end{document}